**Conductive grain systems: relaxation under strong electric fields**

Alexander Z. Patashinski, Mark A. Ratner

Northwestern University, Department of Chemistry, Evanston, IL

*Abstract*

In an external electric field, a system of conductive grains embedded in a dielectric matrix becomes unstable and relaxes towards a conductive state. We describe and discuss the elementary acts of this relaxation. When the grains packing density is large, the relaxation is controlled by narrow gaps separating neighboring grains. The electric fields in some gaps are enhanced, relative to the external field, by the factor $R/\lambda$, $R$ being the grain radius and $\lambda$ the width of the gap. These enhanced fields trigger dielectric breakdown in the gaps. Both breakdown events and grain motions increase the permittivity of the composite; advancement of the breakdown process leads to global failure of insulation. The non-linear and hysteretic charge-discharge behavior of the system is determined by the main parameters characterizing the breakdown, the delay time relative to electric field increase, and the lifetime of breakdown conductivity after the supporting current has vanished.

### *Introduction: conductor-dielectric composites as adaptive media*

Currently, various spatially-heterogeneous systems (composites) are being studied [1-10] as promising materials for energy storage, sensors, memory devices, screening and shielding, and other applications. In particular, embedding electrically conductive grains in an insulating matrix creates a structurally complex system with useful features, especially when the size of a conductive region is in the sub-micrometer range and the grains packing density is close to the limit set by steric attraction of the grains. A common feature of these systems is that an external electric field triggers relaxation processes transforming the system into a conductor. Here, we study elementary events of this relaxation in a system of conductive spheres, and discuss the scenario for propagation of these events. A key role in this scenario belongs to the most narrow gaps between grains, defined by the condition $\lambda << R$ where $R$ is the radius of a sphere and $\lambda$ the width of the gap (the shortest distance between spheres).

A large concentration of narrow gaps is a signature feature of a jammed grain system at or near the steric limit of grain-packing density. When the average packing density is well below this limit, a large concentration of narrow gaps may appear locally due to aggregation creating densely-packed clusters of grains. The steric limit is manifested by a co-continuous cluster of mechanically contacting grains effectively resisting further increase of the packing density. It is important that a mechanical contact means that the steric repulsion is sufficiently strong to prevent further decrease of the distance between grains, but this contact does not imply or exclude large conductivity (electrical contact) between contacting grains. Depending on the fine structure of the grain surface and the grain-matrix interaction, the system at the steric limit may remain insulating or become conductive. A detailed discussion of relations between the steric and the electric contact is beyond the scope of this study. Here, we consider the case when a thin dielectric layer of the width $\lambda_0 << R$ (for example, created by strong binding of the matrix to the grain surface) prevents significant electric conductivity between grains at mechanic contact.

Narrow gaps can be seen as links (bonds) of a near-neighbors (*NN*) network where grains serve as vertices. At or close to the steric limit, this network is co-continuous: in a composite occupying the space between parallel electrodes (see Fig. 1), there are many paths connecting the electrodes. Co-continuity also appears in aggregated composites when the densely packed aggregates percolate. The actual configuration of grains in the system appears as a result of some natural or industrial process, and then, depending on the rheology of the matrix, becomes temporarily or permanently frozen. For example, in a ceramic matrix, the positions of grains are practically permanent. In contrast, in a sufficiently fluid matrix and in polymers, sub-micrometer-sized grains have significant mobility[11-15], so the electrostatic forces between polarized grains may lead to increase in packing density in some regions of the system while decreasing this density in other regions.

Displacements under the action of electrostatic forces represent a form of grain system adaptation to the external electric field. Due to jamming and the resistance of the visco-elastic matrix, these motions are expected to be slow. For grains containing thousands of atoms, the contribution to the free energy $\mathscr{F}$ of the system of the kinetic energy of the grains is negligible. However, the energy of the stress in the matrix caused by grains relative displacements, and the electrostatic energy of the grain system are not negligible. These parts of the free energy depend on temperature, pressure, and on the grain configuration and the slowly relaxing stress in the matrix. One writes then the free energy $\mathscr{F}$ of the system as sum of the electrostatic energy $\mathscr{F}_{es}$ [16,17] and the energy $\mathscr{F}_0$ of the visco-elastic matrix:

$$\mathscr{F}(\varepsilon, E) = \mathscr{F}_0 + \mathscr{F}_{es}, \quad \mathscr{F}_{es} = -V\frac{\varepsilon E^2}{2} \quad . \qquad \qquad \textbf{1}$$

In (1), the dependence of the electrostatic energy $\mathscr{F}_{es}$ on the grains configuration is universally described by the permittivity, $\varepsilon$. Any process further diminishing the (negative) electrostatic energy at $E$=const can be tracked by an increase in $\varepsilon$. Spontaneous relaxation goes in the direction of decreasing the total free energy (1), so spontaneous changes are also controlled by the free energy $\mathscr{F}_0$ of the matrix. The matrix is viscoelastic: at short times, displacements of grains result in elastic stress in the matrix and the corresponding increase (quadratic in displacements [16]) in $\mathscr{F}_0$; at a larger time, this stress relaxes and the elastic energy is dissipated (converted into heat).

A special form of adaptation to electric field is the change in the grains configuration caused by the breakdown of insulation in a gap. In the free energy (1), this event is accounted for by increase in $\varepsilon$. The local breakdown occurs when the electric field in the gap exceeds the electric strength of the matrix. Local electric fields are proportional to the external field $E$ (see next Section). We assume that the applied field is limited by the condition that local breakdown events are rare and appear at large distances from each other. However, the probability of a breakdown in a chosen gap is proportional to the time for which the gap is electrically stressed. Over a long time, accumulation and propagation of breakdown events may lead to total failure of insulation.

Below, a more detailed description and discussion is given of the relaxation process in the grain system subjected to a strong electric field.

### *Electric fields in narrow gaps*

In analytic and numeric studies, a composite system is treated as piecewise medium, where grains are represented by regions of permittivity $\varepsilon_g$ while the embedding matrix has permittivity $\varepsilon_g$. The limit $\varepsilon_g \to \infty$ describes conductive grains in a permanent (or slowly varying) electric field; in this limit, the surface of a grain is equipotential, so each grain $a$ can be characterized by its electrostatic potential $\phi^{(a)}$. For high-permittivity grains ($\varepsilon_g >> \varepsilon_g$) the equipotentiality of a grain surface is an approximation neglecting corrections of the order of ($\varepsilon_m/\varepsilon_g << 1$). Solving the electrostatic problem for a many-body system is a challenging task; for a chosen grain configuration, some results can be obtained numerically (see for example [17-18]). For this study, we need a qualitative, order of magnitude estimate of the largest fields in the matrix for a rather general grains configuration. The electric field in the matrix can be calculated using the method of image charges [16]. Fictitious charges $q(r)$ are placed inside the regions representing the conductors, and the charges and their positions $r$ are found by fitting the conditions on grain surfaces and on electrodes (or at infinity). The electrostatic problem for two conductive spheres ($a$ and $b$) in a uniform external field $E$ has an iterative solution [19, 20]. According to this solution, when the gap between the spheres is narrow ($\lambda/R << 1$), the electric field in the most narrow part of the gap is essentially the field of a dipole made of only two image charges $\pm q$ situated on the line connecting the centers of the spheres (see Fig. 1). The electrostatic potential of this gap dipole is

$$\varphi(r) = \varphi(0) + U^{(ab)} \frac{3R}{4} \left( \frac{1}{\sqrt{r_+^2 + (d+z)^2}} - \frac{1}{\sqrt{r_+^2 + (d-z)^2}} \right) , \qquad \textbf{2}$$
$$r_+^2 = x^2 + y^2, \quad d = \sqrt{3R\lambda}$$

where $U^{(ab)} = \phi^{(a)} - \phi^{(b)}$ is the gap voltage. In the narrow part of the gap (defined by the condition $r_+ << R$), the equipotential surfaces of the dipole are spherical up to corrections that are of higher order in $r_+/R << 1$. In the same approximation, the local electric field in this part is $E^{(ab)} \sim U^{(ab)}/\lambda$. However, to satisfy the boundary conditions outside the narrow part of the gap, one needs to position image charges at distances $\sim R$ from the narrow gaps; these charges (and the electrostatic potentials of the grains) are then self-consistently determined by the entire grains configuration.

For a many grains system, formula (2) gives the electric fields in narrow parts of narrow gaps: when the gap voltage is known, a corresponding gap dipole can be assigned to each narrow gap so that the boundary conditions on grains boundaries are approximately satisfied. A narrow part of a narrow gap serves then as a gap capacitor with surface area $S \approx 2\pi\lambda R$ and the capacity $C = \varepsilon_0 \varepsilon_m S/\lambda$; here, $\varepsilon_0$ is the permittivity of vacuum. For a given gap voltage $U^{(ab)}$ and $\lambda$ decreasing, the capacitor area $S$ is shrinking, but the capacitance $C^{(ab)} = S\varepsilon_0\varepsilon_m/\lambda \sim R\varepsilon_0\varepsilon_m$ remains finite; the electric field $E^{(ab)} = U^{(ab)}/\lambda$ inside the gap capacitor increases, and the image charges on the "electrodes" of the gap capacitor is $q^{(ab)} \sim R\varepsilon_0\varepsilon_m U^{(ab)}$. The electrostatic energy $\delta W = (1/2)C^{(ab)}q^{(ab)2} \sim (\varepsilon_0\varepsilon_m)^3 U^{(ab)2} R^3$ stored in the gap capacitor is of the same order as the electrostatic energy stored outside the narrow part in a region of the size $R$, so the contribution of the gap capacitor to the stored energy (and permittivity) is of the same order as that of the rest of the volume. Then, gap capacitors contribution to permittivity of the composite and to stored energy is proportional to the number of narrow gaps in the system; this contribution is newer dominating but can be essential. In particular, the contribution of the narrow gap capacitors is the energy driving the system to increase the packing density of grains.

One notes that the enhanced electric field may cause a significant electric current $I = SE^{(ab)}/\sigma \approx 2\pi R U^{(ab)}/\sigma$ in the narrow part of the gap. Here, $\sigma$ is the effective (leakage) conductivity of the matrix. The cross-section $S \sim \lambda$ of the narrow part is small, so the current density may be large and lead to local temperature increase; it is not yet known if this factor is essential in initiating a dielectric breakdown.

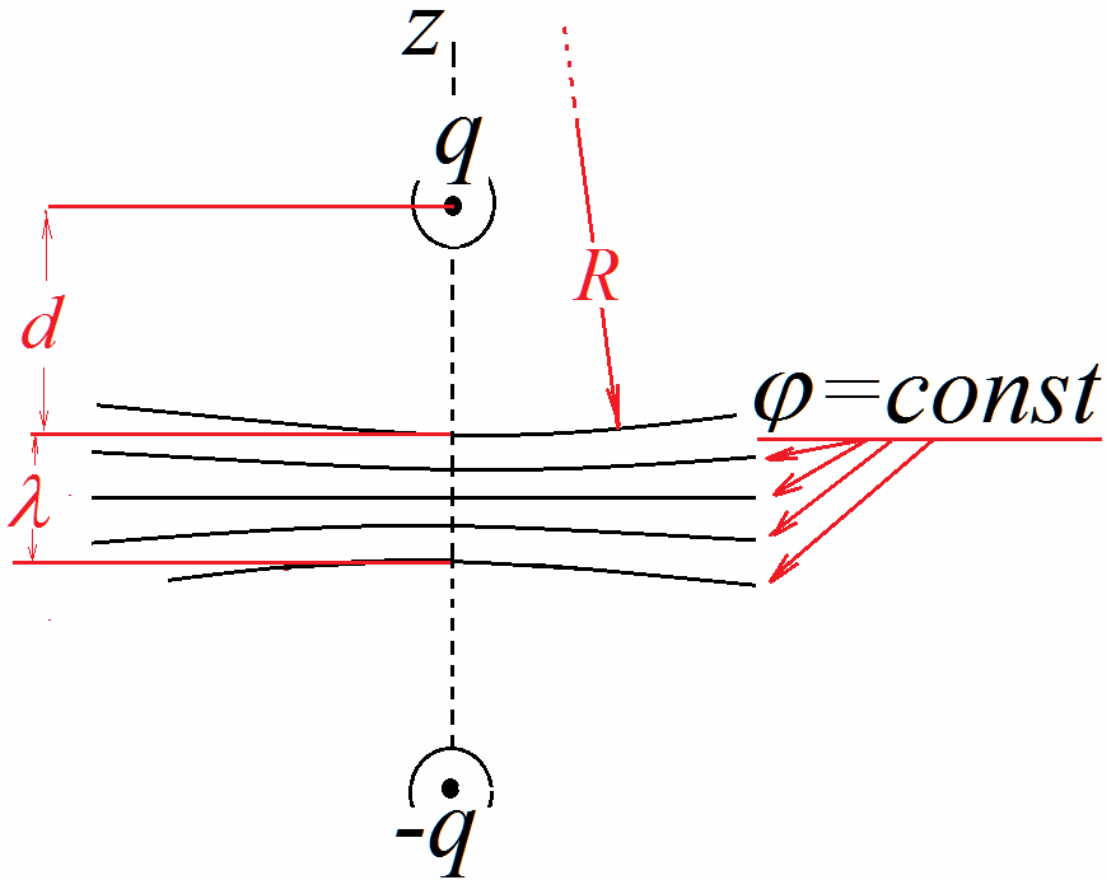

Fig. 1. Equipotential lines in a narrow gap.

According to formula (2), the largest electric fields (and the largest density of electrostatic energy) appear in narrow gaps with the largest gap voltage. We assume that the external electric field is caused by two parallel electrodes (see Fig. 2). To roughly locate the equipotential surfaces, one treats the system as a uniform dielectric medium. In this (effective medium) approximation, the electrostatic potential at a point $r=(x,y,z)$ is $\varphi(r)=\varphi(0)-Ez$, so the equipotential surfaces are parallel to the electrodes. As the next step, a grain $a$ is assigned the effective medium potential $\varphi^{(a)}=\varphi(0)-z^{(a)}E$ of the grain's center $(x^{(a)}, y^{(a)}, z^{(a)})$; in this approximation, the gap voltage between neighboring grains $a$ and $b$ is $U^{(ab)}=\varphi^{a}-\varphi^{(b)}=(z^{(b)}-z^{(a)})E$.

The effective medium approximation describes the electrostatic potential averaged over length-scales where variations of the packing density can be neglected. On smaller length-scales, the granular nature of the system becomes important. In a densely-packed system of grains, the local packing density is approximately constant for regions containing tens or hundreds of grains. In this case, the $NN$-network is co-continuous and contains many short chains connecting the electrodes (blue lines in Fig 2), with each chain consisting of $\sim L/R$ grains linked by gap capacitors. The sum of all gap voltages in the chain is the difference $EL$ between electrodes potentials, so the average gap voltage for the path is $\delta U \sim (ER)$. The above definition of a short chain leaves some freedom to permit chains branching. The system of those short chains is a sub-network of the co-continuous $NN$-network. This sub-network consists of gaps having the largest gap voltages and thus the largest probability of dielectric breakdown in narrow gaps – see next Section.

To conclude this Section, we note that the piecewise medium model assuming the infinitely sharp drop of electric conductivity at the boundary of a conductive grain oversimplifies the physics of conductivity. In real materials, electrons and holes at and near the conductor/dielectric interface are described by Quantum Mechanics that predicts spatial non-locality and charge-carriers tunneling. Another unaccounted for factor is that the surface of a real grain has outstanding asperities, steps and other deviations from a smooth geometrical shape [21].

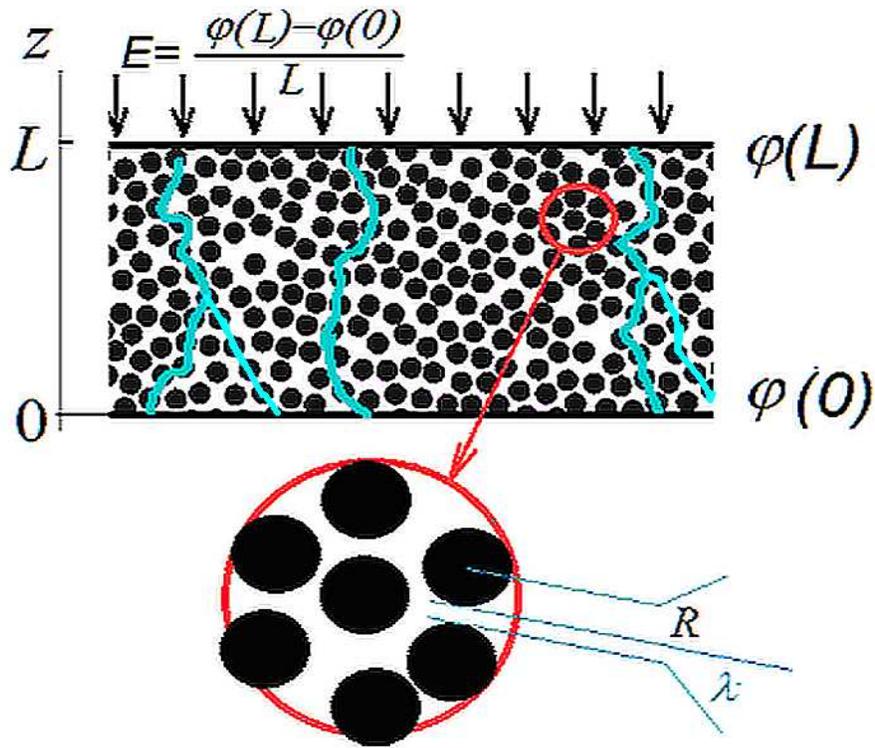

Fig. 2. The grain system in typical application geometry. Blue lines show the short chains of grains connecting the electrodes.

*Breakdown of insulation in thin dielectric layers: a discussion*

When the electric field acting on a thin dielectric layer is small, the leakage current is small; a field of the order or above the electric strength of the dielectric results in a sudden and large increase in the current, a signature of breakdown. Here, sudden means that on the time scale of observations, the time of the transition from the low (leakage) to the large (breakdown) current can be neglected. Soft breakdown in thin ($\lambda$~3-5 nanometers) oxide films in MOS devices has been extensively studied [23-25]; this breakdown is assumed to be triggered by charge traps forming a percolative bridge between electrodes. It is not yet known if this scenario is also applicable to much wider ($\lambda$~10nm-10$\mu m$) liquid or glassy (for example polymeric) or solid (ceramic) layers. Our discussion of the breakdown in these layers is based on available experimental data [2-10, 22, 26-30].

For wide ($\lambda$>10$\mu$m) layers, the electric strength (defined as the electric field causing almost immediate breakdown) increases [9] with decreasing width of the layer, but below some material-specific threshold (~10$\mu m$ for *PMMA*), the electric strength becomes $\lambda$-independent. More accurate measurements reveal that for a stepwise increase of the applied voltage, the jump in the current is always delayed relative to the last step. Repeating experiments give a distribution of delay times [22], with the average delay time $\tau_d(\lambda, U)$ sharply increasing when the electric field decreases below the electric strength.

Breakdown behavior in nano-arrays of golden disks encased in a polymeric (*PDMS, PMMA*) matrix is described in ref [29, 30]; both the disks and the gaps separating the discs are in the tens of nanometers range. These nano-arrays can be seen as two-dimensional (*2D*) nano-composites with a narrow distribution of gap widths. We note that for studying local breakdown, a one-dimensional nano-array (a chain of nano-disks or nano-spheres) has some advantages. The observed sudden increase of current on gradual increase of the applied field is interpreted in [29] as appearance of conductive channels connecting the disks to form conductive pathways between the electrodes. Then, approximate stabilization of the breakdown current indicates that the pathways have large but finite conductance. Once created, the channels remain open for as long as the electric field and the electric current are maintained. However, after the supporting field is turned off, the channels remain conductive for a finite (hours, days) time $\tau_{ch}$, and then the conductivity of the system returns to pre-breakdown level. One notes that the cooling time for a microscopically-narrow channel is microscopically small, so a macroscopic lifetime $\tau_{ch}$ excludes the increase in temperature as the only cause of the large conductivity; one rather expects that channel conductivity is caused by structure and/or chemistry changes.

A conductive channel can be seen as a voltage-controlled conductivity switch. The average delay time $\tau_d(\lambda, U_b)$ for opening this switch after the voltage $U_b$ is turned on, and the average lifetime $\tau_{ch}$ of the post-breakdown conductivity are the most important characteristics of this switch. The most schematic current-voltage ($I$-$U$) diagram for this switch is shown in Fig. 3. Using this diagram, we describe and discuss the expected breakdown behavior in narrow gaps between conductive grains.

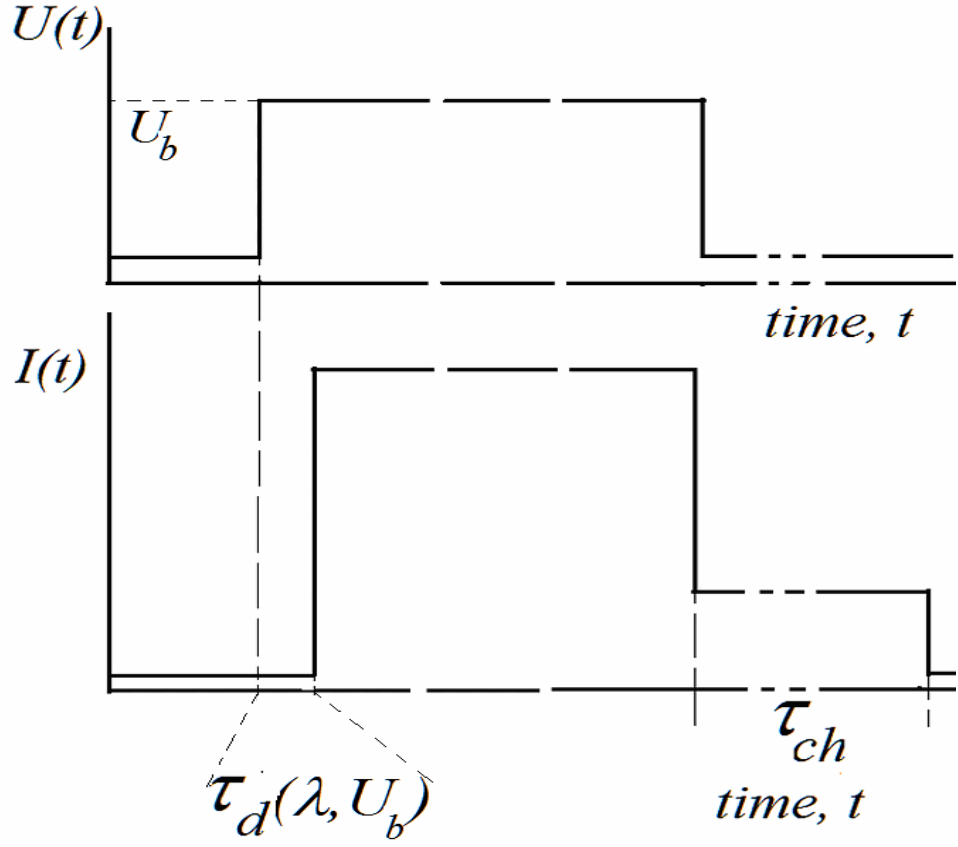

Fig.3. A schematic plot of voltage-current relations in a dielectric layer. See the text for a discussion of breakdown voltage $U_b$, delay time $\tau_d(\lambda, U_b)$, and the lifetime $\tau_{ch}$ of large post-breakdown conductivity.

### Dielectric breakdown in thin gaps between grains

Breakdown of insulation in the gap separating neighboring grains *a* and *b* creates a new conductive grain *ab*, and the initial difference $U^{(ab)}$ between the electrostatic potentials of the grains disappears on times of the order of $\tau_{eq} \sim (C/\mathfrak{R})^{1/2} \sim (\varepsilon_m R/\mathfrak{R})^{1/2}$ where $\mathfrak{R}$ is the conductance of the open channel, and $C \sim R\varepsilon_m$ the capacitance of the new grain. Here, we assume that the lifetime $\tau_{ch}$ of post-breakdown conductivity is sufficiently large: $\tau_{ch} \gg \tau_{eq}$. According to [29], this time in *PDMS* and *PMMA* is of the order of at least many hours. Electrical merging of neighboring grains due to breakdown is an act of adaptation of the system to the electric field. This act decreases the free energy (1), and increases the permittivity of the system. The rate $R_{br}=dN_{br}(t)/dt$ of breakdown events has a strong dependence on the external electric field $E$: this rate is very small for small fields, but becomes very large for strong fields.

The distribution of gap voltages in the system is wide. As described earlier, the largest gap voltages appear between neighboring grains situated on shortest paths between electrodes; these gaps have the largest breakdown probability. Then, the new grains also belong to these paths. The wide gap voltage distribution, and also variations in narrow gap widths determine a very wide distribution of breakdown delay times. The number $N_{ch}(t)$ of conductive channels created by the breakdown process over the time $t$ beginning with the turning on of the external electric field is then expected to be a non-linear function of $t$. At the early stages of this process, the conductive channels appear uncorrelated on large distances from each other. With $N_{ch}(t)$ increasing, redistribution of electrostatic potential due to grains electrical merging increases the probability of breakdown in the gaps separating the newly created grains from neighbors. Then, the rate of breakdown events increases, but now these events become correlated in space and time. This correlation favors creation of chain-like conductive aggregates. At a later time $\tau_p(E)$, these conductive aggregates percolate and make the system a conductor. The increase of the number $N_{ch}(t)$ can be tracked by measuring the time-dependent permittivity of the system.

A more complicated breakdown kinetics appears in the case when the percolation time $\tau_p(E)$ is much larger than the channel lifetime $\tau_{ch}$. The time $\tau_p(E)$ sharply increases when the electric field decreases, so the condition $\tau_p(E) \gg \tau_{ch}$ puts an upper limit on the field $E$. In this case, the number $N_{ch}(t)$ of conductive channels increases on times $t < \tau_{ch}$ but then stabilizes due to the increasing rate $R_{cl} = N_{ch}(t)/\tau_{ch}$ of conductive channels closure. Closure of a conductive channel (*ab*) returns the grains (a) and (b) to an electrically separated state. Due to the presence of electric field $E$, these grains are now oppositely charged and form an electric dipole; the electric field of this dipole compensates the external field to keep the electrostatic potentials of newly separated grains equal. Polarization of the system remains unchanged by channel closing, and continues to increase on times larger that $\tau_{ch}$. continues to increase. Turning off or alternating the external field leaves the system in a temporary ferroelectric state with a non-zero total dipole moment, but creates a difference in the electrostatic potentials of the separated but charged grains. Then, charges on the grains decrease and eventually disappear due to leakage currents or a new breakdown in the gap; the time $\tau_{da}$ of this disappearance depends of leakage conductivity of the gap material. The grains discharge process can be tracked, and the discharge time found by studying the decrease in polarization of the system in zero electric field conditions.

The behavior of the system in an alternating electric field depends on the relations between $\tau_{da}$, $\tau_{ch}$, and the field alternation period $\tau_{alt}$. For short $\tau_{da} \ll \tau_{ch} < \tau_{alt}$, the polarization includes an alternating part with twice the frequency of external field alternation. The electrostatic forces between grains also alternate with that doubled frequency; these forces are known to cause electro-striction [16]. When the time $\tau_{da}$ is of same order as $\tau_{ch}$, the charge-discharge behavior is hysteretic.

*Conclusion*

In a densely-packed system of conductive grains separated by non-conductive gaps, narrow gaps between neighboring grains serve as gap capacitors linking the grains into a co-continuous near-neighbors network. This network significantly increases the electrostatic energy stored in the system under an external electric field, but changes due to local (inside the gap) breakdown events. The system is predicted to show a variety of charge-discharge regimes depending on the relations between the times characterizing the breakdown: breakdown delay time, breakdown conductivity lifetime, conductivity onset time, and internal dipoles discharge time. These times depend on the externally applied field of the system and on properties of the embedding matrix.

It is commonly assumed that the breakdown of insulation in a thin dielectric layer opens a conductive channel. In the simplest model discussed here, a breakdown channel serves as a field-controlled conductivity switch. The nature of the large conductivity in the channel is not yet clearly understood.

We hope that future studies will clarify the mechanisms of initiation and self-organization of the conductive channels, and find a way to engineer the current-voltage characteristic of these channels and a control these characteristics by parameters other than the applied voltage. A recent example of breakdown controlled by a chemical agent can be found in ref. [30].

**Acknowledgements**

AZP acknowledges discussions with Dr. Jiwong Kim of breakdown phenomena in nano-disk arrays. This study was supported by US-Israel Binational grant #60031347.

# References

1. E. Tuncer, Y. V. Serdyuk, S. M. Gubanski, *IEEE Transact.on Dielectrics and Electrical Insulation* **2002**, *9*, 809.

2. M. Kakimoto, A. Takahashi, T. Tsurumi, J. Hao, L. Li, R. Kikuchi, et al. , *Mater. Sci. Eng.* **2006**, B132, 74.

3. Z. M. Dang, Y. H. Lin, C. W. Nan, *Adv. Mater.* **2003**, 15, 1525.

4. Z. M. Dang, Y. Shen, C. W. Nan, *Appl. Phys. Lett.* **2002,** *81*, 4814.

5. L. Qi, B. I. Lee, S. Chen, W. D. Samuels, G. J. Exarhos, *Adv. Mater.* **2005**, *17*, 1777.

6. L. A. Fredin, Z. Li, M. T. Lanagan, et al., *Adv. Funct. Mat.* **2013**, *23***,** 3560.

7. L. A. Fredin, Z. Li, M. T. Lanagan, et al, *ACS Nano* **2013**, *7*, 396.

8. S. A.DiBenedetto, I. Paci, A. Facchetti, et al., *J. of Phys. Chem.* **2006**, *110*, 22394.

9. C. Neusel, G. A. Schneider, *J. Mech. Phys. Solids* **2014**, *63*, 201.

10. B. Wang, X. He, Z., Zhang, et al., *Acc. of Chem. Res.* **2013***, 46*, 761.

11. H. Smaoui, L. E. Mir,  H. Guermazi, et al., *J. of Alloys and Compounds* **2009**, 477, 316.

12. B. Nowack, Th. D. Bucheli, *Envir. Pollution* **2007**, 150, 5.

13. V. V. Ginzburg, S. Balijepailli, *Nano Lett.* **2007**, *7*, 3716.

14. V. V. Ginzburg, K. Myers, S. Malowinski, et al, M<u>acromolecules</u> **2006**, *39*(*11*), 3901.

15. L. D. Landau and E. M. Lifshitz, *Statistical Physics, 3rd ed.*. Elsevir Butterworth-Heinemann Amsterdam-London UK **1980**.

16. L. D. Landau, E. M. Lifshitz, and L. P. Pitaevskii, *Electrodynamics of Continuous Media. 2nd ed.*. Elsevir Butterworth-Heinemann Amsterdam-London UK **1984**.

17. Y. Wu, X. Zhao, F. Li, and Zh. Fan, *J. of Electroceramics* **2003**, *11*, 227.

18. D. A. Robinson, S. P. Friedman, *Physica A: Stat. Mech. and its Appl.* **2005**, *358*, 447.

19. J. Lekner, *J. of Electrostatics* **2011**, *69*, 559.

20. J. Lekner, *J. Appl. Phys.* **2012**, *111*, 076102.

1. B.N.J. Persson, *Sliding friction: physical principles and applications*. Springer. (2000).

2. R. A. Schlitz, K. Yoon, L. A. Fredin, Y.-G. Ha, M. A. Ratner, T. J. Marks, and L. J. Lauhon, *J. Phys. Chem. Lett.* **2010**, 1, 3292–3297.

3. S. Lombardo, J. H. Stathis, B. P. Linder, et al., *J. of Appl. Phys.***2005**, *98*, 121301.

4. J. Sune, E. Y. Wu, S. Tous, *Microelectronic Engineering* **2007**, *84*, 1917.

5. Y.-L. Wu, S.-T. Lin, C.-P. Lee, *IEEE Transactions on device and materials reliability* **2008**, *8*, 352.

6. G. Finis and A. Claudi, *IEEE Transactions on Dielectrics and Electrical Insulation* **2007**, *14*, 487.

7. A. P. Gerratt, and S. Bergbreiter, *J. Micromech. and Microeng.* **2013**, *23*, 67001.

8. J. Huang, S. Shian, I. Roger, M. Diebold, Z. Suo, and David R. Clarke, *Appl. Phys. Lett.* **2012**, *101*, 122905.

9. J. Kim and B. A. Grzybowski, *Adv. Mater.* **2012**, *24*, 1850.

10. E. S. Cho, J. Kim, B. Tejerina, et all, *Nat. Mater.* **2012**, *11*, 978.

**Figure captions:**

Fig. 1. Equipotential lines in a narrow gap.

Fig. 2. The grain system in typical application geometry. Blue lines show the short chains of grains connecting the electrodes.

Fig.3. A schematic plot of voltage-current relations in a dielectric layer. See the text for a discussion of breakdown voltage $U_b$, delay time $\tau_d(\lambda, U_b)$, and the lifetime $\tau_{ch}$ of large post-breakdown conductivity.

**Figures:**

Figure 1

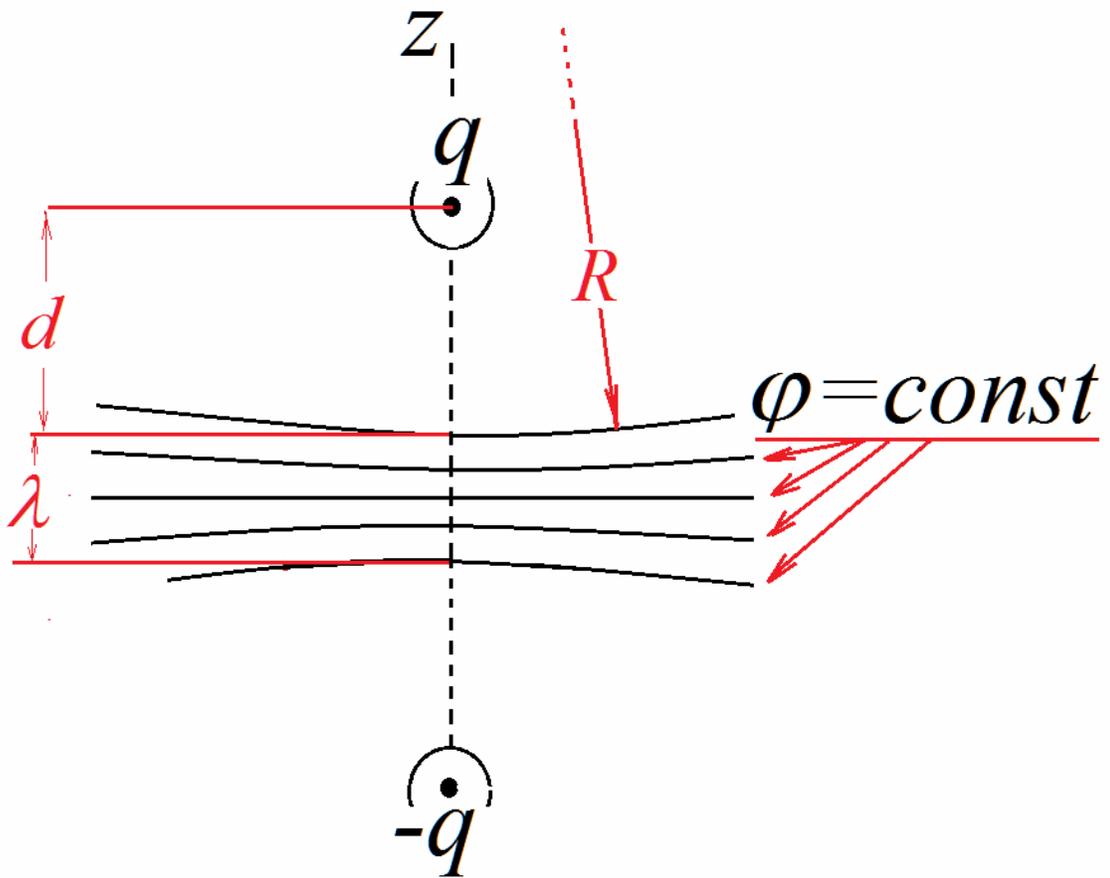

Figure 2

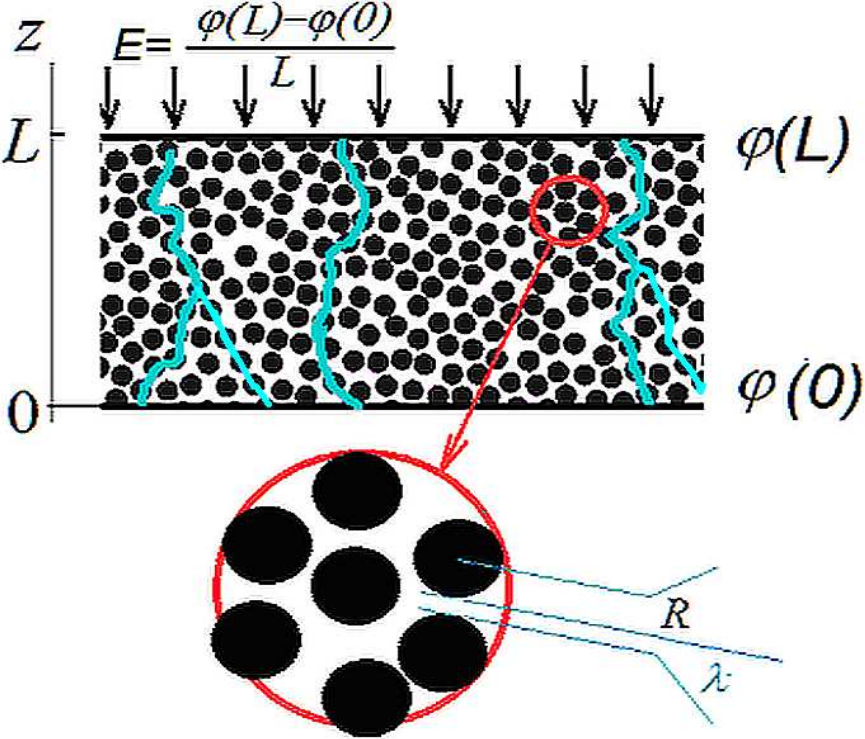

Figure 3

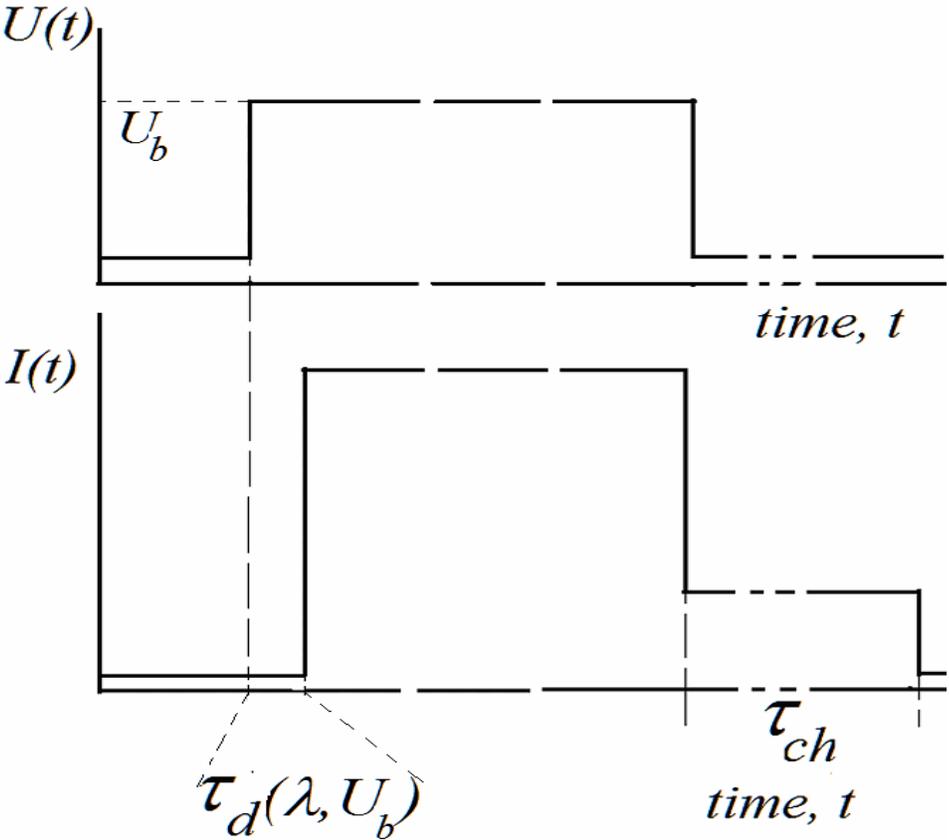